\documentstyle[12pt]{article}
\begin{document}
\newcommand{\newc}{\newcommand}
\newc{\tplanck}{t_{Planck}}
\newc{\mplanck}{m_{Planck}}
\newc{\hnot}{h}

\newc{\be}{\begin{equation}}
\newc{\ee}{\end{equation}}
\newc{\ie}{{\it i.e.}}
\newc{\eg}{{\it eg.}}
\newc{\etc}{{\it etc.}}
\newc{\etal}{{\it et al.}}

\newc{\ra}{\rightarrow}
\newc{\lra}{\leftrightarrow}
\newc{\rhocrit}{\rho_{crit}}
\newc{\rhorad}{\rho_{rad}}
\newc{\rhomatter}{\rho_{matter}}
\newc{\teq}{t_{EQ}}
\newc{\tempeq}{T_{EQ}}
\newc{\trec}{t_{rec}}
\newc{\temprec}{T_{rec}}
\newc{\tdec}{t_{dec}}
\newc{\tempdec}{T_{dec}}
\newc{\abund}{\Omega_0 h^2}
\newc{\abundbaryon}{\Omega_B h^2}
\newc{\deltat}{\delta T}
\newc{\deltatovert}{ {{\delta T}\over T} }
\newc{\deltarho}{\delta\rho}
\newc{\deltarhooverrho}{ {{\delta\rho}\over\rho} }
\newc{\qrms}{\langle Q_{rms}^2\rangle^{1/2}}

\newc{\lsim}{\buildrel{<}\over{\sim}}
\newc{\gsim}{\buildrel{>}\over{\sim}}

\begin{titlepage}
\begin{center}
June 1993\hfill
CfA Preprint No. 3591 \\
             \hfill
\vskip 1in

{\large \bf
The Fourier Space Statistics of Seed-like Cosmological Perturbations
}

\vskip .4in
{\large Leandros Perivolaropoulos}\footnote{E-mail address:
leandros@cfata3.harvard.edu}\\[.15in]

{\em Division of Theoretical Astrophysics\\
Harvard-Smithsonian Center for Astrophysics\\
60 Garden St.\\
Cambridge, Mass. 02138, USA.}
\end{center}

\vskip .2in
\begin{abstract}
\noindent
We propose a new test for distinguishing observationally cosmological models
based on  seed-like primordial
perturbations (like cosmic strings or textures),
from models based on Gaussian fluctuations.
We investigate analytically the {\it Fourier space} statistical
properties
 of temperature or density fluctuation patterns  generated
by seed-like objects and compare these properties with those of Gaussian
fluctuations generated during inflation.
 We show that the proposed statistical test can easily identify
temperature fluctuations produced by a superposition of a small number of seeds
per horizon scale for {\it any} observational angular resolution and {\it any}
seed geometry. However, due to the Central Limit Theorem, the distinction
becomes more difficult as the number of seeds in the fluctuation pattern
increases.

\end{abstract}

\end{titlepage}

\section{\bf Introduction}
\par
One of the directly measurable features of the primordial fluctuations that
gave rise to galaxies and large scale structure formation is the probability
distribution and the corresponding moments of the primordial perturbation
field $\delta (x)$ and its Fourier transform ${\tilde \delta}(k)$.

The phases $\phi_k$ of the Fourier modes ${\tilde \delta}(k)$ are ususally
assumed to be uncorrelated and randomly distributed according to a uniform
probability distribution. This assumption, based on the prediction of
inflationary models, leads by the Central Limit Theorem to a Gaussian
probability distribution for the field $\delta (x)$.

There are two main
advantages of such Gaussian models:
First, all the statistical information about the field $\delta (x)$ is encoded
in a single function: the two point correlation function (or equivalently the
power spectrum). Second, when combined with Cold Dark Matter (CDM), Gaussian
models are in reasonable agreement with small and intermediate scale
observations (White et. al. 1987). However,
observations on large scales (larger than $10h^{-1}$Mpc) have consistently
indicated that Gaussian CDM models lack power on large scales.

One approach to the resolution of this problem is to retain the Gaussian
nature of the primordial perturbations while modifying other aspects of the
model in an effort to transfer power to large scales. This has led to the
construction of the `hydrid models' which attempt through the introduction of
additional parameters (like a component of Hot Dark Matter (HDM)) to reconcile
Gaussian models with large scale structure observations.

The other approach, is the consideration of non-Gaussian primordial
perturbations. A class of non-Gaussian perturbations which is well motivated
physically is seed-like perturbations. These primordial perturbations may be
naturally provided by topological defects (e.g. cosmic strings (Kibble
1976; Vilenkin 1981; Brandenberger 1992)  or textures (Turok 1989))  produced
during phase transitions in the early universe. Other interesting mechanisms
(e.g. primordial black holes (Carr \& Rees 1984)) can also produce seed-like
perturbations.  Models based on cosmic strings for example, have been shown to
have several interesting features that make them worth of further
investigation.  After appropriately normalizing the single free parameter of
the model, cosmic strings  can naturally provide concentrations of galaxies on
sheets (string wakes) with typical dimensions $40\times 40\times 4h^{-1} Mpc^3$
(Vachaspati 1986; Stebbins et. al. 1987; Perivolaropoulos,  Brandenberger \&
Stebbins 1990; Vollick 1992; Hara  \& Miyoshi 1993), they are consistent with
the recent detection of anisotropy by COBE  (Bouchet, Bennett \& Stebbins 1988;
Bennett, Stebbins \& Bouchet 1992;  Perivolaropoulos 1993a), and they are in
reasonable agreement with observations of peculiar velocities regarding
measurements of the Cosmic Mach Number  (Perivolaropoulos \& Vachaspati 1993).
However, like the CDM model,  the cosmic string model is not free from
problems.
As pointed out by Albrecht \& Stebbins (1992a) the
power spectrum of density fluctuations produced by cosmic strings in a universe
consisting mostly of CDM appears to have too much power on small scales. This
problem was shown to be resolved however, if CDM is substituted by HDM
(Albrecht
\& Stebbins 1992b).  In addition, Perivolaropoulos \& Vachaspati (1993) have
recently pointed out that cosmic strings can not explain the observed
magnitude of peculiar velocity flows on scales larger than $50 h^{-1} Mpc$ if
normalized from peculiar velocity observations on scales $5-20 h^{-1}Mpc$. This
problem however, also appears in the CDM model and may be resolved by assuming
velocity bias.

Perturbations in seed-based models may be represented as
a superposition of localized fluctuations with geometry that depends on the
model
under consideration. The most sensitive way to distinguish observationally
models
based on Gaussian perturbations from models based on seeds is by investigating
the statistical properties of the perturbations. In fact, interesting
statistical tests have been proposed that attempt to provide ways to
efficiently
make this distinction
(Coles 1988; Lucchin, Matarrese \& Vittorio 1988;
Scherrer, Melott \& Shandarin 1991; Gaztanaga \& Yokoyama 1992;
Luo \& Schramm 1992; Perivolaropoulos 1993b).

However, there are two main problems that tend to decrease the sensitivity of
these tests. The first comes from the Central Limit Theorem which predicts that
as the number of superimposed seeds increases, the resulting perturbations look
more like Gaussian. The second comes from the finite resolution of
observational experiments. Observations effectively average over patches in the
sky and associate with each patch a measurement that may correspond to a
temperature, a density or a velocity field. By the Central Limit Theorem such
averaging tends to reduce the non-Gaussian signature of seed based models.

The statistical test we discuss in this paper is an attempt to evade the
second problem. By studying the statistical properties of {\it Fourier modes}
we
can isolate the effects of low resolution, to high wavenumber $k$ Fourier modes
and focus on the low $k$ modes that remain unaffected by the smoothing on small
scales.

In what follows we consider perturbation patterns produced by a random
superposition of N identical seed perturbations and derive the probability
distribution and moment generating function of the Fourier modes that
correspond to the pattern.
The pattern of perturbations investigated here is known in the literature as
`shot noise' (Campbell 1909; Rice 1944)
and appears in several and diverse problems. The
statistical properties of shot noise have been studied previously (Rice
1944) mainly in coordinate space
and in the large N limit, showing strong
Gaussian behavior. In the present analysis we focus instead on the statistical
properties in {\it Fourier space}. The results presented here are fairly
general
in that they are valid for any value of N and any shape of the superimposed
seeds.  For simplicity we focus on the one dimensional case but we show that
the
analysis can be easily generalized to higher dimensions.

\section{\bf The Large N Limit}

Consider the random function
\begin{equation}
f(x)=\sum_{n=1}^N f_1 (x-x_n)
\end{equation}
where $f_1 (x)$ is a seed function superimposed randomly at positions $x_n$
such that $-l\leq x_n \leq l$. Both $f(x)$ and $f_1 (x)$ are defined within
the interval $[-l,l]$ and periodic boundary conditions are used during the
superposition. The Fourier expansion of $f_1 (x)$ is:
\begin{equation}
f_1 (x) = \sum_{k=-\infty}^{+\infty} g_1 (k) e^{i {{k \pi}\over l} x}
\end{equation}
with
\begin{equation}
g_1 (k) = {1\over {2l}} \int_{-l}^{+l} dx f_1(x) e^{-i {{k \pi}\over l} x}
\end{equation}
The fact that $f_1 (x)$ is real implies that $g_1 ^\ast (k) =g_1 (-k)$.
Using (1) and (2) the random function $f(x)$ can be expanded as:
\begin{equation}
f(x)=\sum_{k=-\infty}^{+\infty} g_1 (k) e^{i {{k \pi}\over l} x} Q(k)
\end{equation}
with
\begin{equation}
Q(k)=\sum_{n=1}^N e^{-i {{k \pi}\over l} x_n}\equiv q_1 (k;x_1 ..x_N)+i
q_2(k;x_1 ..x_N)
\end{equation}
Thus all the random properties of $f(x)$ have been transferred
to $Q(k)$ which is independent of the shape of the seed function $f_1 (x)$
and can be viewed as the final position of a N step random walk in the two
dimensional $q_1 - q_2$ plane. Notice that the reality condition $Q^\ast(k)
=Q(-k)$ which is trivially satisfied implies that the end points of random
walks with negative $k$ are simply obtained by reflection with respect to the
$q_1$ axis
of the corresponding positive
$k$ end points.
 We are interested in the joint probability
distribution $P (q_1, q_2)$ and the corresponding moment generating function.

 The general case of arbitrary N is treated in the next section. Here we
study the special case $N\rightarrow \infty$ for which there are results
available in the literature (Rice 1944). For $N\rightarrow \infty$ it is
easy to show that the random variables $q_1,q_2$ become independent i.e.
\begin{equation}
P(q_1,q_2)=P(q_1) P(q_2)
\end{equation}
 In the same limit, the Central Limit Theorem implies that both $q_1$
and $q_2$ (being sums of identically distributed random variables) are
distributed according to the Gaussian
\begin{equation}
P(q_i)\rightarrow ({1\over {\sqrt{2 \pi \sigma ^2}}}) e^{-{{(q_i - \mu)^2}\over
{2 \sigma ^2}}}
\end{equation}
where $\mu=<q_i>$ and $\sigma ^2 = <q_i^2>$ ($i=1,2$). Since the probability
distribution of $x_n$ is uniform in the interval $[-l,l]$ we can find
$\mu$ and $\sigma^2$ as:
\begin{equation}
\mu=<q_1>=({1\over {2l}})^N \int_{-l}^{+l} dx_1 ... dx_N (\sum_{n=1}^N \cos
{{{k\pi}\over l}x_n}) = N \delta_{k0}
\end{equation}
and
\begin{equation}
\sigma^2=<q_1^2>={N\over 2}
\end{equation}
Similar results are also easily shown to hold for $q_2$. Thus for $k\neq
0$ we find
\begin{equation}
P(q_1,q_2)\rightarrow {1\over {\pi N}} e^{-{q^2 \over N}}
\end{equation}
where $q^2 \equiv q_1^2 +q_2^2$. Clearly, the probability distribution is
independent of the phase $\phi_k \equiv \tan^{-1} ({{q_1}\over {q_2}})$ of the
Fourier modes. Therefore, for $N\rightarrow \infty$ the Fourier phases are
distributed uniformly while the Fourier mode magnitude $q(k)$ has a Gaussian
distribution. It may be easily seen by visualizing the random walk $Q(k)$ that
the probability distribution of $\phi_k$ will in fact be uniform for any N.
 However, the rest of the results of this section are not valid for
finite N since the independence of the variables $q_1,q_2$ (expressed
through (6)) breaks down in that case. This
will be shown rigorously in the following section.

\section{\bf Arbitrary N}

The Fourier transform of $P(q_1,q_2)$ may be written for any N as:
\begin{equation}
{\bar P}(p_1,p_2)=\int_{-N}^{+N} dq_1 dq_2 P(q_1,q_2)
e^{i{{p_1 \pi}\over N}q_1} e^{i{{p_2 \pi}\over N} q_2}
\end{equation}
which implies that
\begin{equation}
P(q_1,q_2)=({1\over {2N}})^2 \sum_{p_1,p_2 = -\infty}^{+\infty} {\bar
P}(p_1,p_2) e^{-i{{p_1 \pi}\over N}q_1} e^{-i{{p_2 \pi}\over N} q_2}
\end{equation}
where $p_1, p_2$ are integer variables, Fourier conjugate to $q_1,q_2$. By
inspection of (11) it becomes clear that ${\bar P}(p_1,p_2)$ is also the moment
generating function for the distribution $P(q_1,q_2)$. In fact, it is easy to
see that
\begin{equation}
<q_1^m q_2^n>=({{i \pi}\over N})^{-(n+m)} {{\partial^{n+m} {\bar P}
(p_1,p_2)}\over {\partial p_1^m \partial p_2^n}}\vert_{p_1=p_2=0}
\end{equation}
But the same moments are also generated by the function
\begin{equation}
R(p_1,p_2)=({1\over {2l}})^N \int_{-l}^{+l} dx_1 ... dx_N
e^{i{{p_1 \pi}\over N}q_1(k;x_1, ... x_N)} e^{i{{p_2 \pi}\over N} q_2 (k;x_1
... x_N)}
\end{equation}
since ${1\over {2l}}$ is the probability that $x_n$ will be in the range
$[x_n,x_n+dx_n]$. It is easy to check that (13) also holds with ${\bar
P}(p_1,p_2)$ substituted by $R(p_1,p_2)$. Since the moment generating function
that corresponds to a given probability distribution is uniquely defined
(Feller 1971) we must have
\begin{equation}
{\bar P} (p_1,p_2) = R(p_1,p_2)
\end{equation}
By expanding $q_1 (k;x_1 ... x_N)$ and $q_2 (k;x_1 ... x_N)$ according to (5)
we
obtain using (14) and (15)
\begin{equation}
{\bar P}(p_1,p_2)= ({1\over {2 \pi k}} \int_{-k\pi}^{k\pi} d\xi e^{i(t_1 \cos
\xi + t_2 \sin \xi)})^N
\end{equation}
where $t_i \equiv p_i {\pi \over N}$ ($i=1,2$) and $\xi={{k \pi x}\over l}$.
Let now
\begin{eqnarray}
t_1 & = & t \cos \delta \\
t_2 & = & t \sin \delta
\end{eqnarray}
Using the periodicity of $\cos \xi$ and the fact that $k$ is integer (16)
becomes
\begin{equation}
{\bar P} (p_1, p_2) = ({1\over {2\pi}} \int_{-\pi}^{+\pi} d\xi e^{i t \cos
\xi})^N
\end{equation}
or
\begin{equation}
{\bar P}(p_1,p_2)=(J_0 (t))^N
\end{equation}
where
\begin{equation}
t=({\pi \over N})\sqrt{p_1^2 +p_2^2}\equiv ({\pi \over N}) p
\end{equation}
The generating function (20) is one of the central results of this paper. It is
valid for any N and clearly depends only on the magnitude $p$ of the vector
$(p_1,p_2)$. This implies that its Fourier transform
$P(q_1,q_2)$ will similarly be a function of the magnitude $q$ only and there
will be no dependence on the phase of the vector $(q_1,q_2)$. Thus, the Fourier
phases of seed induced perturbations obey a uniform distribution for any number
N of superimposed seeds. Obviously, this statement applies to each mode $k$
individually and does not imply that there will be no correlations among the
phases of different modes. Such correlations will clearly exist for seed
perturbations but are not the subject of the present study.

Our results can explain the numerical simulations of
Suginohara \& Suto  (1991) where it was found that even
in strongly non-Gaussian evolved density fields the phases $\phi_k$ are
uniformly
distributed. The authors of that paper had concluded that the
investigation of the distribution function of the phases $\phi_k$ does not
provide a sensitive test of the non-Gaussian behavior in the strongly
non-linear regime but no clear explanation was given of this fact. Since the
density field in the non-linear regime can be viewed as a superposition of
dense lumps (seeds), the above analysis is applicable and predicts exactly the
uniform distribution of phases seen in the simulations of
Suginohara \& Suto (1991).

The probability distribution $P(q_1,q_2)=P(q)$ is obtained by Fourier
transforming the generating function (20) as follows:
\begin{equation}
P(q)=({1\over {2N}})^2 \sum_{p_1, p_2=-\infty}^{+\infty} (J_0 (t(p_1,p_2)))^N
e^{i{\vec t} \cdot {\vec q}}
\end{equation}
where ${\vec t} ={\pi \over N} (p_1,p_2)$ and ${\vec q}=(q_1,q_2)$. For
$N>1$ we may approximate the sum (22) by an integral and reduce it to
\begin{equation}
P(q)={1\over {2\pi}} \int_0^{\infty} dp\hskip 1mm p (J_0 (p))^N J_0 (pq)
\end{equation}
It is straightforward to verify that for $N>1$
\begin{equation}
\int_{-\infty}^{\infty} dq_1 \int_{-\infty}^{\infty} dq_2 P(q_1,q_2) =
\int_0^{\infty} dp p (J_0 (p))^N \int_0^{\infty} dq q J_0 (pq) = 1
\end{equation}
which is to be expected since $P(q_1,q_2)$ is a probability distribution.

 Directly measurable quantities from a given fluctuation pattern are the
moments of the fluctuation probability distribution.
The moments generated by the
function (20) can easily be obtained and compared with the moments of the
Gaussian probability distribution. Since ${\bar P} (p_1,p_2)$ depends only on
the
magnitude $p$, it is easy to show using (13) that for any positive integers $m$
and $N$ we have \begin{equation} <q_1^m>=<q_2^m>={{\partial ^m (J_0
(t/i))^N}\over {\partial {t}^m}}\vert_{t =0}
\end{equation}
By expanding $(J_0 (t/i))^N$ in powers of $t$ we obtain
\begin{equation}
(J_0 (t/i))^N = 1+{{(t \sqrt{N/2})^2}\over 2} +
{{(t \sqrt{N/2})^4}} ({1\over 8} - {1\over {16N}}) + ...
\end{equation}
{}From (25) and (26) it may be shown that the moments of the appropriatelly
normalized variables $r_i\equiv {q_i \over {\sqrt{N/2}}}$ ($i=1,2$) for $k\neq
0$ are
\begin{eqnarray}
<r_i^{2m+1}> & = & 0 \\
<r_i^2> & = & 1 \\
<r_i^4> & = & 3 (1 - {1\over {2N}})
\end{eqnarray}
The kurtosis (defined as $(<r_i^4> - 3)$) is negative for all finite N
and approaches the Gaussian value 0 for large N. Also, the skewness $<r_i^3>$
is
0 for all N. By expanding the generating function
further, the higher moments may also be obtained. The negative sign of the
kurtosis is to be contrasted with the corresponding sign of the kurtosis of
seed
perturbations {\it in coordinate space}  where several cases of interest have
been shown to have positive kurtosis (Scherrer \& Bertschinger 1991; Luo \&
Schramm 1992; Perivolaropoulos 1993b).

By numerically evaluating the integral (23) we plot the probability
distribution $P(q)$ and compare it with the Gaussian. This is shown in Figure
1 (dotted line) for $N=10$. The corresponding Gaussian distribution with the
same standard deviation (obtained from (10) with $N=10$) is also shown for
comparison (continous line).

It is of interest to obtain the generating
function for the moments of the normalized variables $r_i$. This is easily
shown
to be
\begin{equation}
{\bar P}(t_1, t_2)= (J_0 ({{t}\over {i \sqrt{N/2}}}))^N=
(1+{{{t}^2} \over {2 N}} + ... )^N \longrightarrow e^{{{t}^2 \over 2}}
\end{equation}
where the limit, indicated by the arrow, leading to the standard Gaussian
generating function $e^{{{t}^2 \over 2}}$, is obtained for $N\gg 1$. Thus, the
generating function approaches, as expected, the Gaussian for large N.

Let us demonstrate the utility of these results in a somewhat realistic case.
Consider an one dimensional pattern of fluctuations in Fourier space. In a
realistic case, these fluctuations will be a superposition of a Gaussian
noise random variable $q_n$ and a signal $q_s$. Let the signal to noise ratio
be
\be
\gamma \equiv {{<q_s^2>}\over {<q_n^2>}}
\ee
assumed known. The measured variable at each pixel is $q=q_n + q_s$. We
want to test the hypothesis that $q_s$ is produced by a superposition on $N$
seeds. The moment generating functions for the variables $q_n$ and $q_s$ are:
\be
M_{q_n} (t) = e^{\alpha {t^2\over 2}}
\ee
and
\be
M_{q_s} (t) = (J_0 (t/i))^N
\ee
Therefore $\gamma = {{<q_s^2>}\over {<q_n^2>}} = {N\over {2\alpha}}$. Since the
variables $q_n$ and $q_s$ are independent, the moment generating function for
the measured variable $q$ is:
\be
M_{q_s} (t) = (J_0 (t/i))^N e^{\alpha {t^2\over 2}}
\ee
It is now straightforward to expand  $M_q (t)$ and thus obtain the kurtosis
$\kappa$ for the random variable $q$:
\be
\kappa = {{<q^4>}\over {<q^2>^2}}= 3 (1-{\gamma^2 \over {2(1+\gamma)^2 N}})
\ee
The kurtosis $\kappa$ is measurable, and any constraint on it can be translated
using (35) to a constraint on $N$, the number of superimposed seeds on the
pattern under consideration. Given that different seed-based models predict
widely different number of seeds per Hubble scale (according to simulations,
there are 0.04 textures unwinding per Hubble volume per Hubble time while the
corresponding number for long strings is about 10), this test can be used to
rule in favour of a particular seed-based model or, if $N$ is found too large
to rule out such models. For example, the number of textures predicted to have
unwound in $10\times 10$ degree MBR sky map between the time of recombination
and today is less than 8.
In fact, if reionisation is realized, as required by the texture model
(Turok \& Spergel 1990), this number will be {\it much} less than 8.
 This implies that the proposed test may be
efficiently used in this case since the predicted value of the kurtosis can be
smaller by more than 10\% compared to the Gaussian for $\gamma\simeq 1$.

Expressions similar to (35) may be easily obtained for higher moments of $q$.
Using such expressions the proposed test can be applied even in cases where the
signal to noise ratio $\gamma$ is not known, by using the measured constraints
on
higher moments of $q$ to eliminate $\gamma$.

It is straightforward to generalize the above analysis to higher
dimensional cases. In fact we will show that the form of the generating
function
is the same in any number of dimensions. We will demonstrate the three
dimensional case, applicable to large scale structure considerations.
The two dimensional case corresponding to the MBR follows trivially
from the three dimensional analysis.

Consider a three dimensional rectangular area
with coordinates ${\vec x}=(x_1,x_2,x_3)$ such that $-l\leq x_i\leq l$
($i=1,2,3$). In this case the wavenumber $k$ becomes ${\vec k}=(k_1,k_2,k_3)$
and
using the same analysis as in the one dimensional case it can be shown that
\begin{equation}
{\bar P}(p_1,p_2)=R(p_1,p_2)=(({1\over {2l}})^2
\int_{-l}^{+l} dx_1 \int_{-l}^{+l} dx_2 \int_{-l}^{+l} dx_3
e^{i (t_1 \cos ({{{\vec k} \pi}\over l} {\vec x})+
t_2 \cos ({{{\vec k} \pi}\over l} {\vec x}))})^N
\end{equation}
Define now
\begin{eqnarray}
\xi_{1}&\equiv& {\pi \over l}(k_1 x_1 + k_2 x_2 + k_3 x_3)\\
\xi_{2}&\equiv& {\pi \over l} k_2 x_2 \\
\xi_{3}&\equiv& {\pi \over l} k_3 x_3
\end{eqnarray}
A change of variables from $(x_1,x_2,x_3)$ to ($\xi_1,\xi_2,\xi_3$) leads to
\begin{equation}
{\bar P} (p_1,p_2)= ({1\over {2\pi}})^3 { 1\over {k_1 k_2 k_3}}
\int_{-k_3 \pi}^{+k_3 \pi} d\xi_3 \int_{-k_2 \pi}^{+k_2 \pi} d\xi_2
\int_{-k_1 \pi+\xi_2 +\xi_3}^{+k_1 \pi+\xi_2 +\xi_3} d\xi_1
e^{i(t_1 \cos \xi_1+t_2 \sin \xi_1)}
\end{equation}
which leads to a result identical to the one dimensional result
(19) since $k_1,k_2,k_3$ are integers.
It is trivial to see that the same is true for the two dimensional case.

\section{\bf Discussion-The Power Spectrum}

So far we have investigated the statistical properties of the random function
$Q(k)=q_1 (k) + i q_2 (k)$ which is only part of the Fourier modes of the
perturbations. In fact we are interested in the full mode function
\begin{equation}
g(k)\equiv g_1 (k) Q(k)
\end{equation}
Since the only random part of $g(k)$ is $Q(k)$, the statistical properties of
$g(k)$ are fully specified once we know such properties for $Q(k)$. For
example, the power spectrum of perturbations defined as
\begin{equation}
P(k)=<\vert g(k) \vert ^2>
\end{equation}
is easily found using (29) to be
\begin{equation}
P(k)=\vert g_1 (k) \vert ^2 <q^2(k)>= N \vert g_1 (k) \vert ^2
\end{equation}
for $k\neq 0$ (obviously $P(0)=N^2 \vert g_1 (k) \vert ^2$ since $Q(0)=N$).
It is possible to obtain the same result for the power spectrum by simply
Fourier transforming the two point correlation function in coordinate space
(Rice 1944; Scherrer \& Bertschinger 1991).

In conclusion, we have proposed a new statistical test for the
identification of signatures of seed-based models in cosmological observations.
The main advantages of investigating the statistics of perturbations in Fourier
space rather than in coordinate space is that in Fourier space the
analysis is valid for {\it any} geometry of superimposed seeds and can be
directly applied to any particular experiment by simply selecting the Fourier
modes for which the resolution and sky coverage of the experiment is most
sensitive. No smoothing is needed as would be the case for coordinate space
statistics.

The statistical properties of the seed-like perturbations were shown
to approach the Gaussian for a large number N of superimposed seeds.
Thus, these properties can only distinguish efficiently models where the
perturbations are produced by a small number of seeds per horizon scale. For an
alternative statistic which can efficiently distiguish particular seed based
models for larger N see
Moessner, Perivolaropoulos \& Brandenberger (1993).

 So far we have considered superposition of identical (but of any shape) seeds.
However, our results can be easily generalized to variable
seed magnitude and extend in space. Such generalizations are shown in the
Appendix.

\bigskip

{\bf Acknowledgements}

\noindent
I am grateful R. Brandenberger, R. Moessner and T. Vachaspati for
interesting discussions and for providing heplful comments after
reading the manuscript. I would also like to thank the referee for
particularly helpful comments.
This work was supported by a CfA Postdoctoral Fellowship.

\appendix
\section*{Appendix A}

\leftline{\bf Generalizations}
\vskip 0.1in

An interesting generalization of our results can be provided by
considering seeds of variable size. For example the size of perturbations
induced by topological defects increases with cosmic time due to the growth of
the horizon. Such effect can be taken into account by generalizing the Fourier
space variable $g_1 (k) Q(k)$ to
\begin{equation}
\sum_{j=0}^M g_1 (2^j k) Q(2^j k)
\end{equation}
which corresponds to repeating the superposition of N seeds, M times while
each time modifying the spatial scale of each seed by a factor of
2 in order to take into account the horizon growth (Vachaspati
1992; Perivolaropoulos 1993b;
  Moessner, Perivolaropoulos \& Brandenberger 1993).

In this case the moment generating function ${\bar P}_{sum}
(p_1,p_2)$ for the sum of random variables is given
(Feller 1971) by the product of
the individual generating functions i.e.
\begin{equation}
{\bar P}_{sum} (p_1,p_2)= \prod_{j=0}^M (J_0 (\vert g_1 (2^j k)\vert t))^N
\end{equation}
where the factor $\vert g_1 (2^j k)\vert$
appears because we are now interested in the distribution of the variable (44)
as opposed to simply the variable $Q(k)$.

Finally, it is also straightforward to generalize our analysis to the case of
seeds of variable magnitude. Such generalization would be needed in order to
take into account the variable velocities of long cosmic strings. Consider for
example the superposition of N seeds with Fourier transforms $\lambda_i g_1
(k)$
($i=1, ... , N$) where the coefficients $\lambda_i$ represent the relative
magnitude of seed fluctuations. In this case the Fourier mode $k$ becomes
\begin{equation} g_1 (k) \sum_{n=1}^N \lambda_n e^{i {{k \pi}\over \lambda}
x_n}\equiv g_1 (k) Q_\lambda (k)  \end{equation} and the generating function
for
the variable $Q_\lambda (k)$ is \begin{equation}
{\bar P}_\lambda (p_1,p_2)= \prod_{j=1}^N J_0 ({{\lambda_j t}\over i})
\end{equation}

The above discussion is an
attempt to show that our results are fairly general and can be easily adapted
to the cases of particular seed-based models. Clearly, further work is needed
to adapt the above analysis to any particular model. Work in that direction for
the cosmic string case is currently in progress.
\vskip 1cm

\centerline{\large \bf Figure Captions}

{\bf Figure 1:}  A comparison of the Gaussian (continous line) with $P(q)$ for
$N=10$.
 \newpage
\centerline {\large \bf References}
\vskip .5cm
\begin{flushleft}
Albrecht A. \& Stebbins A. 1992a {\it Phys. Rev. Lett.} {\bf 68}, 2121.
Albrecht A. \& Stebbins A. 1992b {\it Phys. Rev. Lett.} {\bf 69}, 2615.
Bennett D. P., Stebbins A. \& Bouchet F. R. 1992. {\it Ap. J. Lett.}
{\bf 399}, L5.\\
Bouchet F. R., Bennett D. P. \& Stebbins A. 1988. {\it Nature} {\bf
335}, 410.\\
Brandenberger R. 1992. `Topological Defect Models of
Structure Formation  after the COBE discovery of CMB Anisotropies',(Invited
Talk at Erice Course, Sep. 1992)  BROWN-HET-881.\\

Campbell N. 1909. {\it Proc. Cambridge Phil. Soc.} {\bf 15} 117.\\
Carr B. J. \& Rees M. J. 1984. {\it MNRAS} {\bf 206} 801.\\
Coles P. 1988. {\it M.N.R.A.S} {\bf 234}, 509.\\
Feller W. 1971. `An Introduction to Probability Theory
and its Applications', New York: Willey.\\
Gaztanaga E., Yokoyama J. 1992 `Probing the
Statistics of Primordial Fluctuations and its Evolution', Fermilab preprint,
PUB-92-71-A.\\
Hara T. \& Miyoshi S. 1993, Ap. J. {\bf 405}, 419.
Kibble T. W. B. 1976, J.Phys. {\bf A9}, 1387.\\
Lucchin F., Matarrese S. \& Vittorio N. 1988. {\it Ap. J.} {\bf 330}, L21.\\
Luo X., Schramm D. 1992.
`Kurtosis, Skewness and non-Gaussian Cosmological Density Perturbations'
Fermilab
preprint PUB-92-214-A.\\
Moessner R.,  Perivolaropoulos L. \& Brandenberger R., 1993.
{\it Ap. J.} in press.\\
Perivolaropoulos L. 1993a. {\it Phys. Lett.} {\bf B298 }, 305.\\
Perivolaropoulos L. 1993b. {\it Phys. Rev.} {\bf D48}, 1530.\\
Perivolaropoulos L., Brandenberger R. \& Stebbins A. 1990 {\it
Phys. Rev.} {\bf D41}, 1764.\\
Perivolaropoulos L. \& Vachaspati T. 1993. submitted to {\it Ap. J. Lett.}.\\
Rice S. 1944. {\it Bell System Tech J.} {\bf 23} 282.\\
Scherrer R. J. \& Bertschinger E. 1991. {\it Ap. J.} {\bf 381} 349.\\
Scherrer R. J., Melott A. \& Shandarin S. 1991. {\it Ap. J.} {\bf
377} 29.\\
Stebbins A., Veeraraghavan S., Brandenberger R., Silk J. \&
Turok N. 1987 {\it Ap. J.} {\bf 322} 1.\\
Suginohara T. \& Suto Y. 1991. {\it Ap. J.} {\bf 371} 470.\\
Turok N. 1989. {\it Phys. Rev. Lett.} {\bf 63} 2625.\\
Turok N. \& Spergel D. 1990. {\it Phys. Rev. Lett.} {\bf 64} 2736.\\
Vachaspati T. 1986 {\it Phys. Rev. Lett} {\bf 57} 1655.\\
Vachaspati T. 1992. {\it Phys. Lett} {\bf B282}, 305.\\
Vilenkin A. 1981. {\it Phys. Rev. Lett.} {\bf 46} 1169.\\
Vollick D. N. 1992, Ap. J. {\bf 397}, 14\\
White S.,Davis M.,Efstathiou G.,Frenk C. 1987. {\it Nature}
{\bf 330}, 451.
\end{flushleft}
\end{document}